\documentstyle[12pt,amssymb]{article}
\begin{document}

\tolerance=5000

\def\pp{{\, \mid \hskip -1.5mm =}}
\def\cL{{\cal L}}
\def\be{\begin{equation}}
\def\ee{\end{equation}}
\def\bea{\begin{eqnarray}}
\def\eea{\end{eqnarray}}
\def\tr{{\rm tr}\, }
\def\nn{\nonumber \\}
\def\e{{\rm e}}

\begin{titlepage}

\begin{center}
\Large
{\bf The minimal curvature of the universe in modified gravity 
and conformal anomaly resolution of the instabilities}

\vfill

\normalsize

\large{ 
Shin'ichi Nojiri$^\spadesuit$\footnote{Electronic mail: nojiri@nda.ac.jp, 
snojiri@yukawa.kyoto-u.ac.jp} and 
Sergei D. Odintsov$^{\heartsuit\clubsuit}$\footnote{Electronic mail:
 odintsov@ieec.fcr.es Also at TSPU, Tomsk, Russia}}

\normalsize

\vfill

{\em $\spadesuit$ Department of Applied Physics, 
National Defence Academy, \\
Hashirimizu Yokosuka 239-8686, JAPAN}

\ 

{\em $\heartsuit$ Institut d'Estudis Espacials de Catalunya (IEEC), \\
Edifici Nexus, Gran Capit\`a 2-4, 08034 Barcelona, SPAIN}

\ 

{\em $\clubsuit$ Instituci\`o Catalana de Recerca i Estudis 
Avan\c{c}ats (ICREA), Barcelona, SPAIN}

\end{center}

\vfill 

\baselineskip=24pt
\begin{abstract}
We discuss the modified gravity which may produce the current cosmic acceleration of the universe and eliminates the need for dark energy.
It is shown that such models where the action quickly grows 
with the decrease of the curvature define the FRW universe with the 
minimal curvature. It is required the infinite time to reach the minimal curvature during the universe evolution. It is demonstrated that
quantum effects of conformal fields may strongly suppress 
the instabilities discovered in modified gravity. We also briefly speculate 
on the modification of  gravity combined with the presence 
of the cosmological constant dark energy.

\end{abstract}


\end{titlepage}

\noindent
{\bf 1.Introduction.} 
The recent astrophysical data \cite{R, B} indicate to the fact that
current universe
mainly  
consists
of dark energy. Theoretically, it is absolutely dark what this dark energy 
should be, as any of proposed candidates is not free of the internal
problems. Instead of the introduction of some mysterious fluid with large
negative pressure or the  fine-tuning the cosmological constant or
whatever else ... one may
think on the gravitational alternative for the dark energy.
This leads to the modification of the gravity at current universe with the
simplest
possibility: $1/R$ term \cite{CDTT,CCS,v1}. It is natural to expect that
such (or similar)
effective gravity may be produced by string/M-theory/braneworld \cite{sn}.

The study of the (effective) modified gravity with the terms which grow 
when curvature decreases shows that such theory may produce the current 
cosmic acceleration\cite{CDTT}. Unfortunately, it contains the number of
instabilities \cite{Dolgov,Woodard,chiba}. It has been shown \cite{sn1}
that further modification of the modified gravity by $R^2$ or other
higher derivative terms may help to resolve the above instabilities.
 Palatini form of $1/R$ gravity \cite{v,meng,fla} does not contain such
instabilities too and also produces the current cosmic acceleration.
However, new instabilities caused by quantum effects may appear
in Palatini form\cite{fla,wang}. The variants of modified gravity with
$R^2$ or $\ln R$
terms 
have been considered in metric form \cite{sn1,sn2} as well as in Palatini 
version \cite{wang}. Having in mind, that instabilities may be avoided
\cite{sn1, sn2} it could be the right time to think on physical
consequences and further development of modified gravity.

One immediate  conclusion from the form of $1/R$ modified gravity is that 
universe is never minkowskii one. In other words, there is minimal 
curvature of the universe somehow similar to maximal Hagedorn temperature 
of thermal string universe. The infinite time is required to reach 
the singularity point $R=0$ at the universe evolution. It should exist
the mechanism which makes impossible the close approach to the minimal
curvature\footnote{One can suggest the models where Lagrangian is growing 
with the decrease of the curvature but it takes finite value at $R=0$.
The minimal curvature does not appear in such models.}.
Otherwise, the universe becomes highly instable. It could be that 
quantum effects are very important near the minimal curvature.
In other words, one can conjecture the quantum gravity model where some
kind of 
higher derivative gravity is ultraviolet fixed point of the
renormalization group flow. The inflation occurs at this phase.
The intermediate Einstein gravity is some (instable) transitory regime.
Finally, some version of modified gravity with $1/R^n$ is infrared 
fixed point where cosmic acceleration occurs. RG flow should define the
universe evolution and control 
the coefficients of the corresponding effective gravity (fixed points),
as well as the value for the minimal curvature.    
Of course, it is not so easy to develop the quantative picture for above
construction. Nevertheless, one can study some elements towards to 
this direction.

\noindent
{\bf 2.Modified gravity with minimal curvature and cosmic acceleration.} 
Let us start from the quick review of modified gravity formulation.
The  general 4-dimensional action is:
\be
\label{RR1}
S={1 \over \kappa^2}\int d^4 x \sqrt{-g} F(R)\ .
\ee
Here $R$ is the scalar curvature and $F(R)$ is some arbitrary function. 
Introducing the auxiliary fields $A$ and $B$, one may rewrite the
action 
(\ref{RR1}) as following:
\be
\label{RR2}
S={1 \over \kappa^2}\int d^4 x \sqrt{-g} \left\{B\left(R-A\right) + F(A)\right\}\ .
\ee
One may eliminate $B$  and obtain
\be
\label{RR6}
S={1 \over \kappa^2}\int d^4 x \sqrt{-g} \left\{f'(A)\left(R-A\right) + f(A)\right\}\ .
\ee
By using the conformal transformation 
\be
\label{RR7}
g_{\mu\nu}\to \e^\sigma g_{\mu\nu}\ ,
\ee
with 
\be
\label{RR8b}
\sigma = -\ln F'(A)\ ,
\ee
the action (\ref{RR6}) is rewritten as
the Einstein frame action (see also \cite{ms})
\be
\label{RR10}
S_E={1 \over \kappa^2}\int d^4 x \sqrt{-g} \left( R - {3 \over 2}g^{\rho\sigma}
\partial_\rho \sigma \partial_\sigma \sigma - V(\sigma)\right)\ .
\ee
Here
\be
\label{RR11}
V(\sigma)= \e^\sigma G\left(\e^{-\sigma}\right) - \e^{2\sigma} F\left(G\left(\e^{-\sigma}
\right)\right)= {A \over F'(A)} - {F(A) \over F'(A)^2}\ .
\ee
This is the standard form of the scalar-tensor theory where scalar field
is fictitious one.

Now one can put some restrictions to the late time form of the effective
potential. For instance, let us
 assume that when $A$ is small $F(A)$ behaves as
\be
\label{GR1}
F(A)\sim \gamma A^{-n}\left(\ln {A \over \mu^2}\right)^m\ .
\ee
Here we restrict $n$ by $n>-1$ ($m$ is arbitrary).  

Let us assume the FRW metric in the physical (Jordan) frame is:
\be
\label{RD23}
ds^2 = - dt^2 + \hat a(t)^2 \sum_{i,j=1}^3 \hat g_{ij} dx^i dx^j\ .
\ee
In the Einstein frame, the FRW equation looks like
\be
\label{RD24}
3H_E^2 = {3 \over 4}{\dot\sigma}^2 + {1 \over 2}V(\sigma)\ .
\ee
Here we distinguish the quantities in the Einstein frame by the subscript $E$. 
The Hubble parameter $H_E$ is now defined by $H_E\equiv {\dot {\hat a}_E \over {\hat a}_E}$. 
On the other hand, the equation derived by the variation over $\sigma$ is 
\be
\label{RD25}
0=3\left(\ddot\sigma + 3H_E\dot\sigma\right) + V'(\sigma)\ .
\ee
Then when $A$ is small, one obtains \cite{sn2}
\be
\label{GR5}
t\sim t_E^{1 \over n+2}\ ,\quad a_E \sim t_E^{3(n+1)^2 \over (n+2)^2}\ ,
a\sim t^{(n+1)(2n+1) \over n+2}\ .
\ee
This does not depend on $m$. It takes infinite time in order that 
the universe reaches $R=0$. Therefore if we start from positive curvature $R>0$, the universe 
does not go to the region of negative curvature $R<0$.  

Since $H^2={\kappa^2 \over 6}\rho$, the energy density $\rho$ corresponding to 
$\sigma$ may be defined as $\rho={\rho_0 \over t^2}$. 
Here $\rho_0$ is a constant. Denoting the pressure of $\sigma$ by $p$
and substituting the above expressions of $H$ and $\rho$ to the 
conservation law for the energy-momentum tensor
\be
\label{GR5b}
0=\dot \rho + 3H \left( \rho + p \right)\ ,
\ee
we find 
  the effective $w={p \over \rho}$ for  $\sigma$-field 
\be
\label{GR6}
w=-{6n^2 + 7n - 1 \over 3(n+1)(2n+1)}\ .
\ee

 From (\ref{GR5}), the condition that the universe 
could accelerate is ${(n+1)(2n+1) \over n+2}>1$.
Clearly, the effective dark energy $w$ may be within the existing bounds.

It has been mentioned in ref.\cite{chiba} that $1/R$ model which is
equivalent to some scalar-tensor gravity is ruled out as realistic theory 
due to the constraints to such theories.
As the coupling of $\sigma$ with matter is not small \cite{fla}, we now 
investigate the square of scalar mass, which is proportional to$V''(\sigma)$. 

One may consider the case that the present universe corresponds to the 
minimum $A=A_0$ of $V(A)$, where
\be
\label{Rm1}
V'(A_0)=0\ ,\quad V''(A_0)>0\ .
\ee
Then  
\be
\label{Rm2}
\left.{d^2 V(\sigma) \over d\sigma^2}\right|_{A=A_0}=
\left\{\left({d\sigma \over dA}\right)^{-2}
\left.{d^2 V(A) \over dA^2}\right\}\right|_{A=A_0} \ .
\ee
If we adjust the parameters in $F(R)$ so that 
\be
\label{Rm3}
\left.{d\sigma \over dA}\right|_{A=A_0}=0\ ,
\ee
the effective mass of $\sigma$ becomes big and the scalar field $\sigma$  
decouples. Then the obtained theory does not conflict with the cosmological 
observations, say, the solar system observations\cite{will}.


Let us now address the question of minimal curvature of the universe.
To be specific, we start from the variant of the model \cite{CDTT}:
\be
\label{RA1}
S={1 \over \kappa^2}\int d^4 x \sqrt{-g} \left(R - {a \over R - R_0}\right)\ .
\ee
Here $R_0$ is a small constant (minimal curvature). If we assume that the
Ricci tensor $R_{\mu\nu}$ 
is  covariantly constant, the equation of motion corresponding to the action 
(\ref{RA1}) is given by
\be
\label{RA2}
0=R - {2a \over R - R_0} - {a R \over \left( R - R_0 \right)^2}\ .
\ee
The solution of (\ref{RA3}) is given by
\bea
\label{RA3}
R&=& {2 \over 3}R_0 + \alpha_+ + \alpha_-\ ,
\quad {2 \over 3}R_0 + \alpha_+ \zeta + \alpha_- \zeta^2 \ ,
\quad {2 \over 3}R_0 + \alpha_+ \zeta^2 + \alpha_- \zeta\ ,\nn
\zeta&\equiv& \e^{i{2\pi \over 3}} \ ,\nn
\alpha_\pm^3 &=& - {R_0^3 \over 27} \pm i \sqrt{a\left(a^2 
+ {R_0^2 \over 3}a + {R_0^4 \over 27}\right)}\ .
\eea
In case that $R_0^2 \ll a$,  the three solutions in (\ref{RA3}) 
behave as
\be
\label{RA4}
R\sim \pm \sqrt{a} + {2 \over 3}R_0 + {\cal O}\left(R_0^2\right)\ ,
\quad {2 \over 3}R_0 + {\cal O}\left(R_0^2\right)\ ,
\ee
which reproduces the results in \cite{CDTT}. 
On the other hand, if $a\ll R_0^2$
\be
\label{RA4b}
R \sim R_0 \ .
\ee 

After  introducing an auxilliary 
field $A$ 
\be
\label{RA5}
S={1 \over \kappa^2}\int d^4 x \sqrt{-g} \left(
\left(1 + {a \over \left(A - R_0\right)^2}\right)\left(R - A\right)
+ A - {a \over A - R_0}\right)\ ,
\ee
the action (\ref{RA1}) can be further rewritten in a scalar-tensor 
form \cite{chiba} with the potential:
\bea
\label{RA6}
V(\sigma)&=& {a \left( 2A - R_0 \right) \left( A - R_0 \right)^2 \over \left\{
\left(A - R_0 \right)^2 + a \right\}^2 } \ .
\eea
Here a metric tensor $g_{\mu\nu}$ is rescaled by  
\be
\label{RA8}
g_{\mu\nu}\to \e^\sigma g_{\mu\nu}\ .
\ee
Using equations of motion, one can identify $A$ with the scalar
curvature 
$R$ in (\ref{RA1}). 
The potential $V(\sigma)$  (\ref{RA6}) has two maxima when 
\be
\label{RA9}
A= {2 \over 3}R_0 + \alpha_+ + \alpha_-\ ,
\quad {2 \over 3}R_0 + \alpha_+ \zeta^2 + \alpha_- \zeta\ ,
\ee
and a minimum 
\be
\label{RA10}
A = {2 \over 3}R_0 + \alpha_+ \zeta + \alpha_- \zeta^2 \ ,
\ee
which correspond to the solutions in (\ref{RA3}). $V(\sigma)$ has one more 
extra minimum when 
\be
\label{RA11}
A = R_0\ ,
\ee
which appears since the mapping from $A$ to $\sigma$ becomes 
singular when $A=R_0$. For large $A$, 
$V(\sigma)$ behaves as
\be
\label{RA12}
V(\sigma) \sim {2a \over A}\ .
\ee
The potential $V(\sigma)$ vanishes at $A={R_0 \over 2}$ and $A=R_0$. 
The latter corresponds to the minimum in (\ref{RA11}).
However, it is not difficult to show that the infinite time is necessary 
to reach the state with $A=R_0$. Hence, the curvature approaches to the
minimal one but never reaches it. On the same time, the cosmic acceleration naturally occurs in such a model.

\noindent
{\bf 3.Conformal anomaly resolution of the instabilities.} 
Let us consider the role of quantum effects to the FRW universe evolution
and instabilities of above modified gravity.
For simplicity we take into account the quantum effects of conformally 
invariant fields. That may be done by
 adding the trace anomaly induced action $W$ to 
the classical action $S$  (\ref{RA1}):
\bea
\label{RA13}
W&=& b \int d^4x \sqrt{-\widetilde g}\widetilde F \widetilde A \nn
&& + b' \int d^4x \sqrt{-\widetilde g}
\left\{\widetilde A \left[2{\widetilde\square}^2 
+\widetilde R_{\mu\nu}\widetilde\nabla_\mu\widetilde\nabla_\nu 
 - {4 \over 3}\widetilde R \widetilde\square^2 
+ {2 \over 3}(\widetilde\nabla^\mu \widetilde R)\widetilde\nabla_\mu
\right]\widetilde A \right. \nn
&& \left. + \left(\widetilde G - {2 \over 3}\widetilde\square \widetilde R
\right)\widetilde A \right\} \nn
&& -{1 \over 12}\left\{b''+ {2 \over 3}(b + b')\right\}
\int d^4x \sqrt{-\widetilde g} \left[ \widetilde R - 6\widetilde\square \widetilde A 
 - 6 (\widetilde\nabla_\mu \widetilde A)(\widetilde \nabla^\mu \widetilde A)
\right]^2 \ .
\eea 
In (\ref{RA13}), one chooses the metric as 
\be
\label{RA14}
g_{\mu\nu}=\e^{2\widetilde A}\widetilde g_{\mu\nu}
\ee 
and we specify the quantities with $\widetilde g_{\mu\nu}$ by using $\widetilde{\ }$. 
$G$ ($\widetilde G$) and $F$ ($\widetilde F$) are the Gauss-Bonnet
invariant and the square of the Weyl tensor.
\footnote{We use the following curvature conventions:
\begin{eqnarray*}
R&=&g^{\mu\nu}R_{\mu\nu} \\
R_{\mu\nu}&=& R^\lambda_{\ \mu\lambda\nu} \\
R^\lambda_{\ \mu\rho\nu}&=&
-\Gamma^\lambda_{\mu\rho,\nu}
+ \Gamma^\lambda_{\mu\nu,\rho}
- \Gamma^\eta_{\mu\rho}\Gamma^\lambda_{\nu\eta}
+ \Gamma^\eta_{\mu\nu}\Gamma^\lambda_{\rho\eta} \\
\Gamma^\eta_{\mu\lambda}&=&{1 \over 2}g^{\eta\nu}\left(
g_{\mu\nu,\lambda} + g_{\lambda\nu,\mu} - g_{\mu\lambda,\nu} 
\right)\ .
\end{eqnarray*}}

In the effective action (\ref{RA13}), with $N$ scalar, $N_{1/2}$ spinor, $N_1$ vector fields, 
$N_2$ ($=0$ or $1$) gravitons and $N_{\rm HD}$ higher derivative conformal scalars, $b$ 
and $b'$ are given by
\bea
\label{RA16}
&& b={N +6N_{1/2}+12N_1 + 611 N_2 - 8N_{\rm HD} \over 120(4\pi)^2}\nn 
&& b'=-{N+11N_{1/2}+62N_1 + 1411 N_2 -28 N_{\rm HD} \over 360(4\pi)^2}\ .
\eea
In principle, $b''$ may be changed by the finite renormalization of local
counterterm 
in gravitational effective action. Note that such anomaly induced
effective action is the main component of trace anomaly driven inflation
\cite{SMM}. One may choose the metric $\widetilde g_{\mu\nu}$ to be a flat
one: \be
\label{RA17}
\widetilde g_{\mu\nu} dx^\mu dx^\nu = - d\tau^2 + \sum_{i=1,2,3}\left(dx^i\right)^2\ .
\ee
The total action $S+W$ has the following form
\bea
\label{RA18}
S+W &=& {1 \over \kappa^2}\int d^4x \e^{2\widetilde \sigma} \left( R - {a \over R - R_0}\right) \nn
&& + \int d^4 x \left[ {b' \over 2}\widetilde \sigma \widetilde\square^2 \widetilde \sigma 
 - {1 \over 12}\left\{ b'' + {2 \over 3}\left(b+b'\right)\right\}
\e^{2\widetilde \sigma}R^2\right] \ .
\eea
Here $\widetilde \sigma = 2\widetilde A$ and the scalar curvature $R$ is 
given by
\be
\label{RA19}
R=\e^{-\widetilde\sigma}\left( - 3 \widetilde\square \widetilde\sigma - {3 \over 2}
\widetilde g^{\mu\nu} \partial_\mu \widetilde\sigma \partial_\nu \widetilde\sigma \right)\ .
\ee
The equation of motion is
\bea
\label{RA20}
0&=& {1 \over \kappa^2}\left[ \e^{2\widetilde\sigma}\left( R - {3a \over R - R_0} 
 - {aR_0 \over \left(R - R_0 \right)^2}\right) 
+ 3a \widetilde g^{\mu\nu} \partial_\mu \left(\e^{\widetilde\sigma}\partial_\nu\left({1 \over 
\left(R - R_0\right)^2}\right)\right)\right] \nn
&& + b'\widetilde\square^2 \widetilde\sigma + {1 \over 2}\left\{ b'' + {2 \over 3}\left(b+b'\right)
\right\}\widetilde g^{\mu\nu}\partial_\mu \left(\e^{\widetilde\sigma}\partial_\nu R\right)\ .
\eea
One assumes the spacetime is deSitter space
\bea
\label{RA21}
ds^2 &=& {l^2 \over \tau^2}\left( -d\tau^2 + \sum_{i=1,2,3}\left( dx^i \right)^2 \right) \nn
&=& - dt^2 + \e^{{2t \over l}} \sum_{i=1,2,3}\left( dx^i \right)^2 \nn
\tau &=& -l \e^{-{t \over l}} \ ,\nn
R&=&{12 \over l^2}\ .
\eea
Then the equation (\ref{RA20}) has the following form:
\be
\label{RA22}
0={l^4 \over \kappa^2}\left\{ {12 \over l^2} - {3a \over {12 \over l^2} - R_0} 
 - {aR_0 \over \left({12 \over l^2} - R_0\right)^2 } \right\} + 12 b'\ .
\ee
Taking
\be
\label{RA23}
R=R_1\equiv {12 \over l^2}\ , 
\ee
when $b'=0$, Eq.(\ref{RA22}) reproduces Eq.(\ref{RA2}). 
On the other hand, when $a=0$, Eq.(\ref{RA22}) reproduces the anomaly driven 
inflation \cite{SMM}. 

The perturbation from the solution of (\ref{RA22}) is:
\be
\label{RA24}
\widetilde \sigma = \ln {l^2 \over \tau^2} + \delta \sigma\ .
\ee
Here  $l^2$ satisfies eq.(\ref{RA22}). Then by linearizing Eq.(\ref{RA20}) 
and by using Eq.(\ref{RA22}), we obtain
\bea
\label{RA25}
&& 0= {1 \over \kappa^2}\left[ - {24\kappa^2 b' \over \tau^4}\delta \sigma 
+ {l^4 \over \tau^4}\left( 1 + {3a \over \left(R_1 - R_0\right)^2} 
+ {2aR_0 \over \left( R_1 - R_0 \right)^3}\right) \delta R\right] \\
&& + b' \widetilde\square \delta\sigma 
+ \left[ - {6a \over \kappa^2 \left(R_1 - R_0\right)^3} 
+  {1 \over 2}\left\{ b'' + {2 \over 3}\left(b+b'\right)\right\}\right]
\widetilde g^{\mu\nu}\partial_\mu \left({l^2 \over \tau^2}\partial_\nu \delta R\right)\ .
\nonumber
\eea
Here
\be
\label{RA26}
\delta R = - {12 \over l^2}\delta \sigma + {\tau^2 \over l^2}
\left( - 3 \widetilde\square \delta\sigma + {6 \over \tau}\partial_\tau \delta\sigma \right)\ .
\ee
By using the ambiguity of the renormalization for $b''$, we may choose
(as kind of renormalization condition)
\be
\label{RA27}
0=  - {6a \over \kappa^2 \left(R_1 - R_0\right)^3} 
+  {1 \over 2}\left\{ b'' + {2 \over 3}\left(b+b'\right)\right\}\ .
\ee
When $\tau\to 0$, Eq.(\ref{RA26}) behaves as 
\be
\label{RA28}
\delta R \sim - {12 \over l^2}\delta \sigma \ .
\ee
Then the terms inside the first bracket $[\ ]$  (\ref{RA25}) dominate when 
$\tau \to 0$ and
\be
\label{RA29}
\delta \sigma \sim 0\ .
\ee
The limit $\tau\to 0$ corresponds to $t\to +\infty$ in (\ref{RA21}), 
which is the late time (or present time) of the universe. Eq.(\ref{RA29}) tells that the 
perturbation is strongly suppressed. Then the instabilities as in 
\cite{Woodard,fla,chiba} do not occur.
In fact, the quantum effects produce the higher derivative terms which may
suppress the instabilities in the same way as in ref.\cite{sn1}.

 We should stress that, if there is no the quantum 
correction, the theory becomes inconsistent with the present cosmology due to the 
instabilities. 
As $\kappa^{-2}\sim \left(10^{19}{\rm GeV}\right)^2 = \left(10^{28}{\rm eV}\right)^2$, 
if we assume $R_1 \sim R_0 \sim \sqrt{a} \sim \left(10^{-33}{\rm eV}\right)^2$, 
which corresponds to the present acceleration of the universe, we find
\be
\label{RA29b}
b'' + {2 \over 3}\left(b+b'\right) \sim \left(10^{28+33}\right)^2=10^{122}\ ,
\ee
which does not seem to be realistic. One may consider, however, the case
that 
$a\ll R_0^2$ as in (\ref{RA4b}). If $a\sim 10^{-155} {\rm eV}^2$, the parameters 
$b$, $b'$ and $b''$ can be of the order of unity.

In \cite{chiba}, it has been shown that the general modified gravity can be rewritten 
in the form of the scalar-tensor theory. As the order of the coupling of
the extra 
scalar field with matter field is always the same  \cite{fla}, there seems to be a conflict with the solar system 
observations\cite{will} if the scalar field is massless or very light. If the scalar 
field is heavy, however, the scalar field decouples and there is no problem. The 
fluctuation of the scalar field corresponds to $\delta\sigma$ in (\ref{RA24}). Then 
since the scalar field $\delta\sigma$ decouples if the condition (\ref{RA27}) is satisfied, 
which tells that the effective mass of $\delta\sigma$ becomes very heavy, the quantum 
effects  help to avoid the problem with the solar system tests.
Thus, the role of quantum effects is in resolution of the instabilities 
and in making for theory even more difficult the approach to minimal
curvature phase which is highly instable. 

Indeed, what happens when $R\sim R_0$ where the action (\ref{RA1}) becomes 
sigular at $R=R_0$ (minimal curvature when $R$ approaches to it
from above).  Then in (\ref{RA20}), the terms 
including the inverse power of $\delta R = R - R_0$ dominate and we obtain:
\be
\label{RA32}
0\sim - {a R_0 \e^{2\sigma} \over \left(R-R_0\right)^2} - 3a \partial_\tau\left(
{l_0^2 \over \tau^2}\partial_\tau\left({1 \over \left(R-R_0\right)^2}\right)\right)\ .
\ee
Here the length parameter $l_0$ is given by 
\be
\label{RA31}
R_0={12 \over l_0^2}\ .
\ee
Assuming
\be
\label{RA34}
\delta R \sim \alpha \tau^\beta\ ,
\ee
with constants $\alpha$ and $\beta$, one finds 
\be
\label{RA35}
\beta= - {1 \over 2}\ ,\quad -2\ .
\ee
 $\sigma$ may be written as 
\be
\label{RA30}
\tilde\sigma = \ln {l_0^2 \over \tau^2} + \delta\sigma(\tau)\ .
\ee
The first term in (\ref{RA30}) gives the constant curvature of deSitter spacetime 
$R=R_0$. We now assume the second term in (\ref{RA30}) is rather small compared 
with the first term. Then $\delta R$ is given in a  form similar to (\ref{RA26}):
\be
\label{RA33}
\delta R = - {12 \over l_0^2}\delta \sigma + {\tau^2 \over l_0^2}
\left( 3 \partial_\tau^2 \delta\sigma + {6 \over \tau}\partial_\tau \delta\sigma \right)\ .
\ee
 From (\ref{RA34}) and (\ref{RA35}) it follows 
\bea
\label{RA36}
\delta \sigma \sim \alpha' \tau^\beta\ ,\quad &&\beta= - {1 \over 2}\ ,\quad -2 \nn
&& \alpha'={l_0^2 \alpha \over 3\left(\beta^2 + \beta - 4\right)}\ .
\eea
Since $\delta\tau$ and $\delta R$ become large when $\tau\to 0$, the above approximation 
breaks down for large $t$. As the cosmological time $t$ is given by (\ref{RA21}), 
$R\sim R_0$ is approximately solution when $t\to -\infty$ but the perturbation grows up 
with $t$. 
If one changes the direction of the $t$, we may obtain a solution where
the curvature $R$ approaches to the critical value $R_0$, 
although it takes infinite time in order for $R$ to reach $R_0$. 
In such a model, as the spacetime is 
asymptotically deSitter, we have effective cosmological constant and we find 
that $w\sim 1$. Here $w$ is the ratio of the effective pressure $p$ and the effective 
energy density $\rho$: $w={p \over \rho}$. 
$\delta\sigma$ in (\ref{RA36}) will give 
a correction to $w$. Including the contribution from $\delta\sigma$, since 
$dt=\e^{\sigma \over 2}d\tau$, the cosmological time $t$ is modified as
\be
\label{RA37}
t\sim -l\left(\ln \tau + {\alpha' \over 2\beta}\tau^\beta\right)\ ,
\ee
or
\be
\label{RA38}
\tau \sim \e^{-{t \over l}}\left( 1 - {\alpha'(1+\beta) \over 2\beta}
\e^{-{\beta t \over l}}\right)\ .
\ee
For FRW metric: $ds^2 = - dt^2 + a^2(t) \sum_{i=1,2,3}
\left(dx^i\right)^2$, we find
\be
\label{RA39}
a\sim l\e^{t \over l}\left( 1 + {\alpha' (1+\beta) \over 2\beta}\e^{-{\beta t \over l}}
\right)\ .
\ee
Since the Hubble constant is given by $H={\dot a \over a}$ and from the FRW equations  
 $H={\kappa^2 \over 6}\rho$ and $\dot H = - {\kappa^2 \over 4}(\rho + p)$, we find
\be
\label{RA40}
\rho \sim {6 \over \kappa^2 l^2}\left( 1 - \alpha' l \left(1 + \beta\right)
\e^{-{\beta \over l}t}\right)\ ,\quad 
p\sim -\rho - {6 \over \kappa^2 l^2}{\alpha' l \beta\left(1 + \beta\right) \over 3}
\e^{-{\beta \over l}t} \ .
\ee
we find
\be
\label{RA41}
w={p \over \rho}=- \left(1 + {\alpha' l \beta\left(1 + \beta\right) \over 3}
\e^{-{\beta \over l}t} \right)\ .
\ee
Since $\beta$ is negative, the correction from $w=-1$ is growing
with the time. 

\noindent
{\bf 4.Discussion.}  In summary, it is shown that recently suggested 
models of modified gravity contain the minimal curvature.
In other words, the universe is never flat one.
The infinite time is necessary in order to reach the minimal curvature 
in the current epoch of cosmic acceleration.
It is remarkable that quantum effects of conformal fields 
may help in stabilizing of such gravity.

Many more variants of modified gravity may be suggested.
For instance, one  may combine the modification of the gravitational action 
at the present epoch with the presence of the cosmological constant (also 
kind of dark energy).
The simplest example of such theory which admits current cosmic speed-up and has no the minimal curvature is given below:
\be
\label{RA42}
F(A)={\e^{\beta A} - \alpha \over \beta}
\ee
Here we assume $\beta>0$. Then 
\be
\label{RA43}
\sigma = - \beta A = - \beta R\ ,
\ee
and
\bea
\label{RA44}
V(\sigma)&=& \left(A - {1 \over \beta}\right)\e^{-\beta A} + {\alpha \over \beta}
\e^{-2\beta A} \nn
&=& -{\left(\sigma + 1\right)\e^\sigma \over \beta} + {\alpha \over \beta}\e^{2\sigma}\ .
\eea
If 
\be
\label{RA45}
\alpha> {\e \over 2}\ ,
\ee
the potential $V$ is  monotonically decreasing (increasing) function of $A$ ($\sigma$). 
If the condition (\ref{RA45}) is satisfied, there is no  
 deSitter space solution. The FRW equation (\ref{RD24}) and the 
$\sigma$-equation (\ref{RD25}) can be solved as a power series:
\bea
\label{RA46}
\sigma &=& \sigma_1 t_E + \left(\mp {3\sigma_1 \over 2}\sqrt{{\sigma_1^2 \over 4} 
+ {\alpha - 1 \over 6\beta}} - {\alpha - 1 \over 3\beta}\right)t_E^2 
+ {\cal O}\left(t_E^3\right) \ ,\nn
H_E &=& \pm \sqrt{{\sigma_1^2 \over 4} + {\alpha - 1 \over 6\beta}} 
+ \left( - {3 \over 4} \sigma_1 \pm {\left(\alpha -1\right)\left(\sigma_1 - 1\right) 
\over 6\beta \sqrt{{\sigma_1^2 \over 4} + {\alpha - 1 \over 6\beta}}}\right)t_E 
+ {\cal O}\left(t_E^2\right) \ .
\eea
Here $\sigma_1$ is a constant of integration (one more constant of the integration can be 
absorbed into the constant shift of $t_E$). We may assume $\sigma_1<0$ so that 
$R>0$ in (\ref{RA43}). As a result
\be
\label{RA47}
t \sim \e^{{1 \over 2}\sigma_1 t_E}\ ,\quad 
a \sim \e^{\left(\sigma_1 + \mp \sqrt{{\sigma_1^2 \over 4} + {\alpha - 1 \over 6\beta}}
\right) t_E} \sim t^{1 \mp \sqrt{ 1 + {2\left(\alpha -1\right) \over 3\beta \sigma_1^2}}}\ .
\ee
Then if $\alpha>1$ (or more strongly Eq.(\ref{RA45}) is satisfied), the universe is 
accelerating for the branch with $+$-sign in (\ref{RA47}). 
Note that the rate of the acceleration is dynamically determined by the 
constant of integration $\sigma_1$.  The effective $w$ is found as
\be
\label{RA48}
w=-1 +{2 \over 3\left(1+x\right)}\ ,\quad 
x\equiv  \sqrt{ 1 + {2\left(\alpha -1\right) \over 3\beta \sigma_1^2}}\ .
\ee
$w$ is a monotonically decreasing function of $x$. When $x\to 1$, which is the limit of 
$\beta\to \infty$ or $\sigma_1 \to \infty$,  $w\to - {1 \over 3}$. 
On the other hand, when $x\to +\infty$, which is the limit of 
$\beta\to 0$ or $\sigma_1 \to 0$, we obtain $w\to - 1$. The equation of state satisfies
\be
\label{RA49}
-1<w<-{1 \over 3}\ ,
\ee
which may be consistent with the present observational constraint 
$-1.45<w<-0.74$.  The above example indicates that true dark energy may 
be the combination of the (weak) gravitational alternative
 for dark energy, cosmological constant and quantum models of the sort 
discussed in \cite{parker, elizalde}.

\noindent
{\bf Acknowledgments} 
The research is supported in part by the Ministry of
Education, Science, Sports and Culture of Japan under the grant n.13135208
(S.N.), DGI/SGPI (Spain) project BFM2000-0810 (S.D.O.), RFBR grant 03-01-00105
(S.D.O.) and LRSS grant 1252.2003.2 (S.D.O.).

\end{document}